\documentclass{article}

\begin{document}

\title{Gravitomagnetic Clock Effects in Black Hole Spacetimes}

\author{
Donato Bini\\
Istituto per Applicazioni della Matematica C.N.R., \\
I--80131 Napoli, Italy \\
and I.C.R.A. International Center for Relativistic Astrophysics, \\
University of Rome, I--00185 Rome, Italy\\ 
\\
Robert T. Jantzen\\
Department of Mathematical Sciences, Villanova University,\\ 
Villanova, PA 19085, USA\\
and I.C.R.A. International Center for Relativistic Astrophysics, \\
University of Rome, I--00185 Rome, Italy
}

\maketitle

\begin{abstract}
Gravitomagnetic clock effects for circularly rotating orbits 
in black hole spacetimes are studied from a relative observer 
point of view, clarifying the roles played by special observer families.  
\end{abstract}

PACS: 0420C

\section*{Preface}

One of us (R.J.) knew Ruggiero through our mutual friend Giovanni
Platania for several decades and was happy that finally, with a
collaborator (D.B.) relocated to Napoli, it appeared that we could begin
to interact more directly both socially and professionally. Indeed we
all had a wonderful dinner together in the summer of 2000. This
potential was cut short prematurely, saddening both of us tremendously.
In our many years of crossing paths, Ruggiero was always such a
gentleman, with obvious class, and had a good sense of optimism and
humor and a special Neopolitan way of looking at the world that we will
never forget. He continues to live on in our hearts and memories with
special affection.

\section{Introduction}

Recently, various clock effects in stationary axially symmetric
spacetimes have renewed interest in better understanding them and
possibly eventually measuring some of them. It is precisely the presence
of a rotating source of the gravitational field that leads to these
effects, first in black hole spacetimes which have the most interest as
the exterior fields of physically interesting time independent axially symmetric sources, and
secondarily in other members of the larger symmetry class to which they
belong, including the G\"odel and van Stockum spacetimes, which provide
alternative examples of what can happen in theory. Rotating sources thus mean
nontrivial gravitomagnetism;
the Sagnac effect, synchronization gaps, and other clock comparison
effects all fall (or should fall) under this umbrella topic referred to
as ``gravitomagnetic clock effects." In fact, one can study them all in
parallel and their relationships using a general approach set in the
context of stationary axisymmetry. 

It is important to consider the observers who are associated with each
of these effects, since most of the effects are observer-dependent.
Gravitoelectromagnetism is basically a long word for a way of describing
spacetimes in terms of families of test observers, and so it appropriate
to use its related tools in approaching the present questions.
Each such family measures spacetime quantities relative to its own local
space and time directions, thus performing a ``relative observer
analysis." The quotient of the spacetime by the family of world lines of
the observer family is useful for considering and making sense of the
``closed circular loops" that we normally think of when visualizing
these effects. In fact for stationary circular world lines, it is really
only the geometry of the intrinsically flat timelike cylinder containing them which is
relevant, where the observer splittings are easy to picture and the
local splitting of each tangent space extends to the whole cylinder by
the symmetry, with global complications which are at the heart of the
various clock effects.

One must distinguish three distinct gravitomagnetic clock effects for a 
pair of oppositely rotating circular geodesic test particles (oppositely 
rotating with respect to an intermediate observer) (see \cite{bijama} and references therein):
\begin{enumerate}
\item
the observer-dependent single-clock clock effect: the difference between 
the periods of the oppositely rotating geodesic test particles as measured 
by the observer's clock \cite{idcf2},
\item
the observer-dependent two-clock clock effect: the difference between the 
periods of the oppositely rotating geodesic test particles as measured by 
their own clocks for one revolution with respect to an observer
\cite{cohmas,bonste,sem}, and 
\item
the observer-independent two-clock clock effect: the difference between the 
periods of the oppositely rotating geodesic test particles as measured by their 
own clocks between two crossing events
\cite{mitpul,fermas,tart1,tart2}.
\end{enumerate}

In the first two cases, for a given observer, one compares the periods of one 
revolution of these orbits (starting from and returning to the same observer 
world line) measured either by the observer's own clock (single-clock effect) 
or by the clocks carried along the two orbits (two-clock effect). 
In the third case, no observer enters the calculation.

The Sagnac effect is similar but with a pair of oppositely rotating 
accelerated photons replacing the geodesic test particle pair. 
The observer-dependent single-clock effect in this new context is the Sagnac effect.
The desynchronization effect is similar but with accelerated spacelike 
curves (orthogonal to the observer family) replacing the null curves 
of the Sagnac effect. This observer-dependent single-clock effect, 
divided by 2, is the synchronization gap which measures the failure 
to synchronize the observer family proper times around a single loop.  

Here we draw some connections between the various clock effects, the 
Sagnac and desynchronization effects, and the usual symmetry adapted 
coordinates in  black hole spacetimes, while extending some previous work. 

\section{Metric splitting, observer-adapted frames,\newline 
     lapse/shift notation}

An orthogonally transitive stationary axially symmetric line element 
is usually written in the  Boyer-Lindquist-like 
coordinates $\{t,r,\theta,\phi \}$ adapted to the timelike Killing observers:
\begin{eqnarray}
ds^2 &=& ds^2_{(t,\phi)}+ds^2_{(r, \theta)}\ ,
\end{eqnarray}
where
\begin{eqnarray}
ds^2_{(t,\phi)}&=& g_{tt} dt^2 + 2 g_{t\phi} dt d\phi+ g_{\phi\phi} d\phi^2\ ,\nonumber \\
ds^2_{(r, \theta)}&=& g_{rr}dr^2+g_{\theta\theta} d\theta^2;
\end{eqnarray}
Decomposing the $t,\phi$ 2-metric into perfect square form 
defines the lapse ($M$), shift ($M_\phi$) and spatial metric 
quantity ($\gamma_{\phi\phi}$) in the so-called threading point of view of the Killing observers
\begin{equation}
ds^2_{(r, \theta)}
= -M^2 (dt - M_\phi d\phi)^2 + \gamma_{\phi\phi} d\phi^2\ .
\end{equation}

The 4-velocities of circularly rotating timelike 
orbits/observers (i.e. at fixed $r$ and $\theta$) form a 1-parameter 
family most simply parametrized by their coordinate ``angular velocity" $\zeta$
\begin{equation}
\label{U}
U=\Gamma (\partial_t+\zeta \partial_\phi) \ ,
\end{equation} 
where $\Gamma = dt/d\tau_U  >0$ is a normalization factor defined by
\begin{equation}
\Gamma =  \{-[g_{tt} + 2 \zeta g_{t\phi} + \zeta^2 g_{\phi\phi}]\}^{-1/2}\ .
\end{equation} 
In order for $U$ to be timelike, $\zeta$ must belong to the interval
$(\zeta_-,\zeta_+)$ between the roots of the quadratic equation
$\Gamma^{-2}=0$ in $\zeta$ corresponding to the two null directions, namely 
\begin{eqnarray}
\label{eq:zetapm}
 \zeta_\pm 
   &=& [-g_{t\phi} 
           \pm (g_{t\phi}{}^2 - g_{\phi\phi}g_{tt})^{1/2}]
                /g_{\phi\phi}  \ .
\end{eqnarray}
(Note $\zeta_-\leq0\leq \zeta_+$.)
Thus apart for an arbitrary normalization factor $\Gamma_{\rm(null)}$,
the photon 4-momentum vector $ P_\pm $ is of the same form as $U$
\begin{equation}
 P_\pm 
    = \Gamma_{\rm(null)}
          ( \partial_t + \zeta_\pm \partial_\phi ) \ .
\end{equation}

We assume that the source of the rotating gravitational field 
is rotating in the positive sense, i.e.\ in the increasing $\phi$ direction. 
Corotating and counter-rotating orbits will refer to motion in the positive ($\zeta>0$) 
and negative ($\zeta<0$) $\phi$ direction respectively in this coordinate system.
For example, the 4-velocities of the corotating ($+$) and counter-rotating ($-$) circularly rotating timelike geodesics are
\begin{equation}
\label{Upm}
U_\pm=\Gamma_\pm (\partial_t+ \dot\phi_\pm \partial_\phi) \ ,
\end{equation}
where it is assumed that $\zeta_-< \dot\phi_- <0< \dot\phi_+ < \zeta^+$.
These represent the orbits of freely falling test particles.

It is convenient to introduce new coordinates adapted to a generic circularly 
rotating stationary observer $U$
\begin{equation}
\tilde t =t, \qquad \tilde\phi = \phi -\zeta t \ , 
\end{equation}
so that the new form of the metric is
\begin{equation}
ds^2_{(\tilde t,\tilde\phi)}
= g_{\tilde t \tilde t} d\tilde t^2 
+ 2 g_{\tilde t \tilde \phi} d\tilde t d\tilde \phi
+ g_{\tilde \phi \tilde \phi} d\tilde \phi^2
=ds^2_{(t,\phi)}\ ,
\end{equation}
or equivalently
\begin{equation}
ds^2_{(\tilde r, \tilde \theta)}
= -\tilde M{}^2 (dt - \tilde M_{\tilde \phi} d\tilde\phi)^2 
  + \tilde\gamma_{\tilde\phi \tilde\phi} d\tilde\phi{}^2\ .
\end{equation}
The 4-velocity of the observer $U$ can be represented in 
exactly the same form as with respect to the original $(t, \phi)$ coordinates
\begin{equation}
U = \tilde \Gamma [\partial_{\tilde t}+\tilde \zeta \partial_{\tilde \phi}]
  = [-g_{\tilde t \tilde t}]^{-1/2} \partial_{\tilde t} \ ,
\end{equation}
where $\tilde \zeta=\zeta-\zeta=0$ and 
$\tilde \Gamma=(-g_{\tilde t \tilde t})^{-1/2}$.
Finally, the direction $\bar U$ orthogonal to $U$ has the form Eq.~(\ref{U}) with the replacement 
\begin{equation}
\zeta \to \bar \zeta = -\frac{g_{tt}+\zeta g_{t\phi }}{g_{t\phi}+\zeta g_{\phi\phi}}.
\end{equation}
The quantity $\bar\zeta$ is an inverse angular velocity measuring the local effective rate of change of the synchronization gap with respect to angle.

Table 1 summarizes the representations of the new observer 4-velocity 
and unit 1-form and their orthogonal counterparts.
The new lapse, shift and spatial metric quantities are explicitly
\begin{eqnarray}
\tilde M ^2 
&=&-g_{\tilde t \tilde t}
=-(g_{tt}+2\zeta g_{t\phi}+\zeta^2 g_{\phi\phi}) 
=\Gamma^{-2}\ , \nonumber \\
\tilde M_{\tilde \phi }  
&=&-g_{\tilde t \tilde \phi} / g_{\tilde t \tilde t}
=\Gamma^2 (g_{t\phi+\zeta g_{\phi\phi}})
=(\bar \zeta -\zeta )^{-1} \ ,\nonumber \\
\tilde\gamma_{\tilde \phi \tilde \phi}
&=& g_{\tilde \phi \tilde \phi}+ \tilde M ^2 \tilde M_{\tilde \phi }^2
=\gamma_{\phi\phi} g_{tt}/ g_{\tilde t \tilde t}\ .
\end{eqnarray}

The coordinate time $t=\tilde t$ can be used to parametrize the world 
lines of the observer $U$ or any other timelike or null world line. This 
may be converted to a proper time with respect that observer differentially by
\begin{equation}
d\tau_U =\Gamma^{-1} dt =\tilde M d\tilde t\ .
\end{equation}
Integrating this along a world line segment gives the equivalent proper time 
elapsed along an observer world line between the beginning and ending time hypersurface.

\typeout{********* TABLE 1}
%
\typeout{*** Table 1. (t-\phi  \tilde t -\tilde \phi)}
\begin{table}[h]\footnotesize
\caption
{New stationary observer quantities in terms of old and new adapted coordinates.}
\vbox{%
 \def\hline{}
 \typeout{*** eqnarray struts inserted}
 \def\Strut{\relax\hbox{\vrule width0pt height 10.5pt depth 5.5pt}}
\begin{eqnarray*}
\begin{array}{lll} 
\hline \Strut
& (t,\phi ) & (\tilde t , \tilde \phi)\\ 
\hline \Strut
U    
& =\Gamma [\partial_t +\zeta \partial_\phi ]
& =\tilde M^{-1}\partial_{\tilde t}    \\ 
\hline \Strut
U^\flat    
& =\Gamma (g_{tt}+\zeta g_{t\phi}) [dt-1/{\bar \zeta}\, d\phi ]
& =-\tilde M[d\tilde t -\tilde M_{\tilde \phi}d\tilde \phi ] \\ 
\hline \Strut
\bar U 
& =\bar \Gamma [\partial_t +\bar \zeta \partial_\phi ] 
& =(\tilde \gamma_{\tilde\phi \tilde\phi})^{-1/2}
[\partial_{\tilde \phi}+\tilde M_{\tilde \phi}\partial_{\tilde t}] \\ 
\hline \Strut
\bar U^\flat 
& =-\Gamma^2/{\bar \Gamma} \, (g_{t\phi}+\zeta g_{\phi\phi}) \zeta 
   [dt-1/{\zeta} \, d\phi]
& =(\tilde \gamma_{\tilde \phi \tilde \phi})^{1/2} d\tilde \phi \\
\hline
\end{array}
\end{eqnarray*}
}
\label{tab:1}
\end{table}

\section{Clock effects}

Consider a pair of oppositely rotating orbits, $U_1$ and $U_2$, as 
seen by an intermediate observer $U$,  with angular 
velocities $\zeta_1$, $\zeta_2$ and $\zeta$ respectively with respect to the original 
coordinates $(t,\phi)$ satisfying $\zeta_1<\zeta <\zeta_2$.
We have in mind identifying $U_{1,2}$ either with the oppositely 
rotating geodesic pair (counter-rotating: $U_1=U_-$, 
corotating: $U_2=U_+$) or with oppositely rotating 
photons (counter-rotating: $U_1=P_-$, corotating: $U_2=P_+$) and 
new coordinates $(\tilde t,\tilde \phi)$ will be adapted to the observer $U$
as described in the previous section.
The unit tangent vector along the orthogonal spacelike counterpart 
of timelike rotating curves $U_{1,2}$ (i.e. corresponding to the 
orthogonal direction in the $t-\phi$ space, oriented in the sense 
of increasing $\phi$) will be denoted by  $\bar U_{1,2}$.

We now consider each of the various clock effects in turn, analyzed 
from the relative-observer point of view.

\subsection{Timelike observer-dependent single-clock clock effect}

The timelike single-clock clock effect relative to the observer $U$ 
measures the observer proper time elapsed between the arrivals of the 
two geodesics (one co-rotating and the other counter-rotating) which 
depart from the same event on that observer worldline (i.e. the 
geodesics each make a single spacetime loop beginning and ending 
at the observer worldline). 
Denoting by $C_{U_\pm}^+$ the line with unit tangent 
vector $U_\pm$ oriented according the direction of increasing time,
this single-clock proper time difference is just the 
difference of the observer periods of the two loops

\begin{eqnarray}\label{eq:1-c}
\Delta \tau_{\rm (1-c)} (U,U_-,U_+)
&=& \int_{C_{U_+}^+} d\tau_U - \int_{C_{U_-}^+} d\tau_U \nonumber \\
&=& \frac{1}{\Gamma} [ \int_{C_{U_+}^+} d\tilde t -\int_{C_{U_-}^+} d\tilde t ] \nonumber \\
&=&\frac{2\pi}{\Gamma}[\frac{1}{\dot\phi_+-\zeta}+\frac{1}{\dot\phi_--\zeta}]\nonumber \\
&=&-\frac{4\pi}{\Gamma}\frac{\zeta -\zeta_{\rm (gmp)}}{(\zeta-\dot\phi_-)(\zeta - \dot\phi_+)}\ ,
\end{eqnarray}
where $\zeta_{\rm (gmp)}=(\dot\phi_++\dot\phi_-)/2$ is the 
angular velocity of the `geodesic meeting point observers' \cite{mitpul,fermas}, 
whose world lines pass through the successive alternating 
meeting points of each pair of corotating and counter-rotating 
geodesics, corresponding to undergoing complete revolutions with respect to those observers.
 
The computation of these integrals is  conveniently done by 
changing the integration variable from $\tilde t$ to $\tilde \phi$ and noting the following facts
a) along the world lines $C_{U_\pm}^+$ the 1-form 
$\bar U_\pm\propto d\tilde t -\frac{1}{\zeta -\dot\phi_\pm} d\tilde \phi$ 
vanishes identically; 
b) along $C_{U_-}^+$ (counter-rotating), $\tilde\phi$ goes from 
$0$ to $-2\pi$ while along $C_{U_+}^+$ (co-rotating), $\tilde\phi$ goes from $0$ to $2\pi$.

As a function of the observer angular velocity $\zeta$, the single clock effect
$\Delta \tau_{\rm (1-c)} (U,U_-,U_+)$ is 
positive for $\zeta>\zeta_{\rm (gmp)}$ (which means the 
corotating geodesic returns later than the counter-rotating one), 
zero corresponding to $U=U_{\rm (gmp)}$ 
and negative for $\zeta<\zeta_{\rm (gmp)}$, 
and it is  a monotonically increasing function of $\zeta$, 
from $\zeta =\dot \phi_-$
(where a vertical asymptote is located) 
to $\zeta =\dot \phi_-$ (where another vertical asymptote is located).

\subsection{Lightlike observer-dependent single-clock clock effect or Sagnac effect}

The lightlike single-clock clock effect relative to the 
observer $U$ \cite{pos,ash,hen,andbilste,ste1,ste2} is a 
measure of the observer proper time elapsed between the 
arrivals of the two photons (one co-rotating and the other 
counter-rotating) at a particular observer, having both departed 
from a single event on that observer's world line (i.e. the 
photons each make a single spacetime loop around the observer $U$). 
Denoting by $C_{P_\pm}^+$ the photon world line with null tangent 
vector $P_\pm$ oriented according the direction of increasing time
this single-clock proper time difference is just the Sagnac effect
\begin{eqnarray}\label{eq:sagnac}
\Delta \tau_{\rm (1-c)} (U,P_-,P_+)
&=&  \int_{C_{P_+}^+} d\tau_U -\int_{C_{P_-}^+} d\tau_U\nonumber \\
&=& \frac{1}{\Gamma} [ \int_{C_{P_+}^+} d\tilde t-\int_{C_{P_-}^+} d\tilde t] \nonumber \\
&=&\frac{2\pi}{\Gamma}[\frac{1}{\zeta_+-\zeta}+\frac{1}{\zeta_--\zeta}]\nonumber \\
&=&-\frac{4\pi}{\Gamma}\frac{\zeta -\zeta_{\rm (nmp)}}{(\zeta-\zeta_-)(\zeta-\zeta_+)}\ ,
\end{eqnarray}
where $\zeta_{\rm (nmp)}=(\zeta_++\zeta_-)/2$ is the angular velocity of 
the `null meeting point observers', 
whose world lines contains the alternating successive meeting points of the 
two corotating and counter-rotating photons after making  complete revolutions 
with respect to those observers. These observers are also 
called `slicing observers' because their with 4-velocity $n$ 
is orthogonal to the $t=constant$ hypersurfaces which form a slicing of the spacetime itself
$\zeta_{\rm (sl)}=\zeta_{\rm (nmp)}$.

The Sagnac effect $\Delta \tau_{\rm (1-c)} (U,P_-,P_+)$ vanishes 
when $U=U_{\rm (nmp)}=n$ and its behavior as a function of $\zeta$ 
is similar to that of $\Delta \tau_{\rm (1-c)} (U,U_-,U_+)$ but the 
two vertical asymptotes correspond now to the photon angular velocities.

One can study the relations that might exist between the timelike 
single-clock clock effect and its lightlike analogue. Recently  an 
observer family has been found~\cite{bijama} for which
\begin{equation}
\Delta \tau_{\rm (1-c)} (U,U_-,U_+) = \Delta \tau_{\rm (1-c)} (U,P_-,P_+)\ .
\end{equation}
In other words the Sagnac effect and 1-clock clock effect agree.
These observers turn out to be the so-called `extremely accelerated observers' \cite{def,sem0,page}. 
Their 4-velocity is related to those of the co- and 
counter-rotating geodesics by a renormalized average
\begin{equation}
U_{\rm (ext)}=\frac{U_+ + U_-}{||U_+ + U_-||} \ ,
\end{equation} 
and their angular velocity is therefore:
\begin{equation}
\zeta_{\rm (ext)}
=\frac{\Gamma_+ \dot\phi_++ \Gamma_- \dot\phi_-}{\Gamma_++ \Gamma_-}\ .
\end{equation}
The most important additional properties of these observers are briefly summarized here:
\begin{enumerate}
\item They see the two geodesics (co-rotating and counter-rotating) with equal magnitude but opposite sign relative velocities:
\begin{equation}
\nu(U_+, U_{\rm (ext)})=-\nu(U_-, U_{\rm (ext)});
\end{equation}
\item Their  4-acceleration is extremal among the whole family of circular orbits parametrized by the angular velocity:
\begin{equation}
\partial_\zeta a(U) |_{\zeta =\zeta_{\rm (ext)}}=0;
\end{equation}
\item They are ``intrinsically non-rotating". In fact the first 
and second torsion of the spacetime Frenet-Serret frame associated 
with them are both vanishing. Consequently the spacetime Frenet-Serret 
angular velocity is also vanishing and the (intrinsic) spatial frame is 
Fermi-Walker dragged along their own world lines. Thus if the extremely 
accelerated observers carry a test gyroscope along with them,  there is  
no gyroscope precession when referred to their intrinsic Frenet-Serret axes.
\end{enumerate}

\subsection{Observer-dependent two-clock clock effect}

The two-clock clock effect with respect to a generic 
observer $U$ is defined as the difference between the proper
periods of two geodesic loops (one co-rotating and the 
other counter-rotating) with respect to the observer $U$:

\begin{eqnarray}
\Delta \tau_{\rm (2-c)} (U,U_-,U_+)
&=&  \int_{C_{U_+}^+} d\tau_{U_+} - \int_{C_{U_-}^+} d\tau_{U_-}\nonumber \\
&=&   \frac{1}{\Gamma_+}\int_{C_{U_+}^+} d\tilde t - \frac{1}{\Gamma_-}\int_{C_{U_-}^+} d\tilde t\nonumber \\
&=&\frac{2\pi}{\Gamma_+  (\dot\phi_+-\zeta)}+\frac{2\pi}{\Gamma_- (\dot\phi_--\zeta)}\nonumber \\
&=& 2\pi \frac{\Gamma_+ +\Gamma_-}{\Gamma_+ \Gamma_-}\frac{\zeta-\zeta_{\rm (ext)}}{(\zeta-\dot\phi_-)(\dot\phi_+-\zeta)}\ ,
\end{eqnarray}

This shows another important property which must be added 
to those enumerated in the previous subsection for the extremely accelerated observers:
\begin{equation}
\Delta \tau_{\rm (2-c)} (U_{\rm (ext)},U_-,U_+)=0\ .
\end{equation}
Their observer-dependent 2-clock clock effect is zero.

\subsection{Observer-independent two-clock clock effect}

The observer-independent two-clock clock effect coincides 
with the observer-dependent two-clock clock effect with 
respect to the geodesic meeting point observers $U_{\rm (gmp)}$,
whose worldline contains the alternating successive meeting 
points of the two oppositely rotating geodesics:

\begin{eqnarray}
\Delta \tau_{\rm (2-c)} (U_{\rm (gmp)},U_-,U_+)
&=&2\pi \frac{\Gamma_+ +\Gamma_-}{\Gamma_+ \Gamma_-}\frac{\zeta_{\rm (gmp)}- \zeta_{\rm (ext)}}{(\zeta_{\rm (gmp)}-\dot\phi_-)(\dot\phi_+-\zeta_{\rm (gmp)})}\nonumber \\
&=& 4\pi \frac{\Gamma_- -\Gamma_+}{\Gamma_+ \Gamma_-} \frac{1}{\dot\phi_+ -\dot\phi_-}\ .
\end{eqnarray}
This effect is observer-independent since it only involves the meeting 
points of the two geodesics after a full revolution (during which they meet twice after the initial departure event).

\subsection{Synchronization gap}

The amount of observer proper time $\tau_U$ elapsing before 
the arrival of the orthogonal spatial orbit with unit tangent 
vector $\bar U$ is what has been called the synchronization gap \cite{baz}

\begin{equation}
\Delta \tau_{\rm (SG)}(U,\bar U)
=\int_{C_{\bar U}^+}d\tau_U
= \frac{2\pi }{\Gamma}\frac{1}{ (\bar\zeta- \zeta)}\ .
\end{equation}

It is nonzero for observer families with nonzero vorticity, 
which prevents the synchronization of the proper times on the 
whole family of observer world lines by an orthogonal slicing. 
Orthogonal synchronization can always be carried out by curves orthogonal 
to the observer world lines, but if they complete a loop around the 
symmetry axis, the initial and final times are desynchronized, hence a `desynchronization effect' occurs.
The synchronization gap is  a measure of this effect. It  vanishes  
for the slicing observers ($\bar \zeta_{\rm (sl)}\to \infty$) and is just half the Sagnac effect:
\begin{equation}
\Delta \tau_{\rm (SG)}(U,\bar U)
= \frac12 \Delta \tau_{\rm (1-c)} (U,P_-,P_+) \ .
\end{equation}

\section{Kerr black holes}

We conclude by specializing our results to the 
equatorial plane of the Kerr spacetime.
The $t$-$\phi$ 2-metric in Boyer-Lindquist coordinates is
\begin{equation}
ds^2_{(t,\phi)}= (-1+2{\cal M}/r)dt^2-4a{\cal M}/r dt d\phi +(r^2+a^2+2a^2{\cal M}/r)d\phi^2\ ,
\end{equation}
and the geodesic angular velocities are
\begin{equation}
\dot\phi_\pm ^{-1} = a\pm (r^3/{\cal M})^{1/2}\ .
\end{equation}
In the limit $r\to \infty$ (which corresponds to a 
realistic situation for experiments in the solar system) the coordinate gamma factors reduce to
\begin{equation}
\Gamma_\pm 
\simeq 1+\frac{3{\cal M}}{2r}+\frac{27{\cal M}^2}{8r^2}\mp 3 \frac{a}{r}(\frac{{\cal M}}{r})^{3/2}\ ,
\end{equation}
so that

\begin{eqnarray}
\zeta_{\rm (nmp)} &\simeq & -2\frac{a{\cal M}}{r^3}\ , \nonumber \\
\zeta_{\rm (gmp)} &\simeq & -\frac{a{\cal M}}{r^3}\ , \nonumber \\
\zeta_{\rm (ext)} &\simeq & -\frac{a{\cal M}}{r^3}(1+3\frac{\cal M}{r})\ . \nonumber \\
\end{eqnarray}

Denoting by $m$ the 4-velocity of the static 
observers (with $\zeta=0$) one finds
\begin{eqnarray}
\Delta \tau_{(1-c)}(m,U_-,U_+) 
&\simeq & 4\pi a\  ,\nonumber \\
\Delta \tau_{(1-c)}(U_{\rm (ext)},U_-,U_+) 
&\simeq & -\frac{12\pi a{\cal M}}{r}(1+\frac{2{\cal M}}{r})\  ,\nonumber \\
\Delta \tau_{(2-c)}(m,U_-,U_+) 
&\simeq & 4\pi a (1+\frac{3{\cal M}}{2r})\  ,\nonumber \\
\Delta \tau_{(2-c)}(U_{\rm (gmp)},U_-,U_+) 
&\simeq & \frac{12\pi a{\cal M}}{r}(1+\frac{3{\cal M}}{2r})\  . 
\end{eqnarray}

\section{The Carter observers}

In black hole spacetimes, the 4-velocity $u_{\rm(car)}$ of the Carter observers \cite{carter} in the equatorial plane is tied to those of the usual threading observers ($m$) and the circular geodesics ($U_\pm$) by the following pair of parallel relationships at the global and local levels
\begin{eqnarray}
\label{1}
\int_{C_{\bar u_{\rm (car)}}}dt 
&=& \frac12 [\int_{C_{U_+}}dt-\int_{C_{U_-}}dt]\ , \nonumber \\
\frac{1}{\bar \zeta_{\rm (car)}}
&=&\frac12 [\frac{1}{\dot\phi_+}+\frac{1}{\dot\phi_-}]\ ,
\end{eqnarray}
where $\zeta_{\rm(car)}=a/(r^2+a^2)$, $\bar\zeta_{\rm(car)}=1/a$. 
The right hand side of the first equation is half the coordinate single-clock clock effect for the threading observers (compare with Eq.~(\ref{eq:1-c})), while its left hand side is the corresponding coordinate synchronization gap for the Carter observers, where the loop is defined with respect to the threading observers (compare with Eq.~(\ref{eq:sagnac})).

This is in direct analogy with the same pair of corresponding equations for the photon orbits and the threading observer Sagnac effect
\begin{eqnarray}
\label{2}
\int_{C_{\bar m}}dt 
&=& \frac12 [\int_{C_{P_+}}dt-\int_{C_{P_-}}dt] \ ,\nonumber \\
\frac{1}{\bar \zeta_{\rm (thd)}}
&=&\frac12 [\frac{1}{\zeta_+}+\frac{1}{\zeta_-}]\ .
\end{eqnarray}

These reciprocal averaging relationships are complementary to the direct averaging relationships 
$\zeta_{\rm(gmp)}=(\dot\phi_+ + \dot\phi_-)/2$, 
$\zeta_{\rm(sl)}=\zeta_{\rm(nmp)}=(\zeta_+ + \zeta_-)/2$. 
All four such relationships may be restated in terms of averaging relationships among the coordinate-time renormalized 4-velocities and corresponding 1-forms and their orthogonal counterparts (i.e. the coefficient of $\partial_t$ or $dt$ is 1).
In contrast, the extremely accelerated observers are defined by an averaging relationship for the unit 4-velocities.

\section{Conclusions}

The various gravitomagnetic clock effects in 
stationary axially symmetric spacetimes have been discussed from a relative observer point of view, including the closely related Sagnac and desynchronization effects. 
The analysis has shown the special roles played by the `geodesic meeting point observers' 
and the `extremely accelerated observers', which have been previously introduced in the literature
in a different context (test particle motion, relativistic definition of inertial forces). 
The discussion is also relevant in view of possible solar system experiments to test 
the validity of the theory of general relativity.

\section*{Acknowledgment}

We thank Bahram Mashhoon for our mutual 
collaboration in revisiting this topic, leading to the 
clarification of our understanding of the larger picture into which all of these effects fit.

\end{document}